\documentclass[aps,prl,twocolumn]{revtex4}
\usepackage{amssymb}
\usepackage{amsmath}

\usepackage{graphicx}

\begin{document}

\title{Chemical potential oscillations from a single nodal pocket in the underdoped high-$T_{\rm c}$ superconductor YBa$_2$Cu$_3$O$_{6+x}$}
\author{Suchitra~E.~Sebastian$^1$}
\email{suchitra@phy.cam.ac.uk}
\author{N.~Harrison$^2$}
\email{nharrison@lanl.gov}
\author{M.~M.~Altarawneh$^2$}
\author{Ruixing Liang$^{3,4}$}
\author{D.~A.~Bonn$^{3,4}$}
\author{W.~N.~Hardy$^{3,4}$}
\author{G.~G.~Lonzarich$^1$}
\affiliation{
$^1$Cavendish Laboratory, Cambridge University, JJ Thomson Avenue, Cambridge CB3~OHE, U.K\\
$^2$National High Magnetic Field Laboratory, LANL, Los Alamos, NM 87545\\
$^3$Department of Physics and Astronomy, University of British Columbia, Vancouver V6T 1Z4, Canada\\
$^4$Canadian Institute for Advanced Research, Toronto M5G 1Z8, Canada}
\date{\today}

\begin{abstract}
The mystery of the normal state in the underdoped cuprates has deepened with the use of newer and complementary experimental probes. While photoemission studies have revealed solely `Fermi arcs' centered on nodal points in the Brillouin zone at which holes aggregate upon doping, more recent quantum oscillation experiments have been interpreted in terms of an ambipolar Fermi surface, that includes sections containing electron carriers located at the antinodal region. To address the question of whether an ambipolar Fermi surface truly exists, here we utilize measurements of the second harmonic quantum oscillations, which reveal that the amplitude of these oscillations arises mainly from oscillations in the chemical potential, providing crucial information on the nature of the Fermi surface in underdoped YBa$_2$Cu$_3$O$_{6+\textrm x}$.  In particular, the detailed relationship between the second harmonic amplitude and the fundamental amplitude of the quantum oscillations leads us to the conclusion that there exists only a single underlying quasi-two dimensional Fermi surface pocket giving rise to the multiple frequency components observed via the effects of warping, bilayer splitting and magnetic breakdown.  A range of studies suggest that the pocket is most likely associated with states near the nodal region of the Brillouin zone of underdoped YBa$_2$Cu$_3$O$_{6+\textrm x}$ at high magnetic fields.
\end{abstract}

\maketitle

While high-$T_{\rm c}$ cuprate superconductors have a history spanning a quarter of a century~\cite{bednorz1,bonn1}, quantum oscillations in these materials have only been measured as recently as a few years ago~\cite{doiron1}. Interestingly, their advent has signalled a re-evaluation of the electronic structure in the normal state induced by high magnetic fields in the underdoped cuprates~\cite{leboeuf1, jaudet1, sebastian2, audouard1, sebastian1, sebastian3, ramshaw2, millis1, chakravarty1}. Where previously, the very existence of quasiparticles was questioned in the normal state of the underdoped cuprates~\cite{lee1}, quantum oscillations have uncovered quasiparticles consistent with the existence of a Fermi liquid within this regime~\cite{sebastian1}.

Despite the progress made by quantum oscillations that complement a broad array of other experiments in discerning the normal state of the cuprates, fundamental mysteries remain both in interpreting the results of quantum oscillation measurements, and in reconciling them with other experimental techniques. Quantum oscillation results have been interpreted in terms of an ambipolar Fermi surface with multiple pockets~\cite{leboeuf1, sebastian2, sebastian3, millis1, chakravarty1} (i.e.) comprising hole carriers concentrated at the nodes ${\bf k}_{\rm nodal}=[\frac{\pi}{2a},\frac{\pi}{2a}]$ and electron carriers located near the antinodes ${\bf k}_{\rm antinodal}=[{\frac{\pi}{a}},0]$ \& $[0,{\frac{\pi}{a}}]$, where $a$ is the lattice constant. This picture, however, is in direct contrast with the observation made by photoemission experiments of a significant ($\approx$ 50 meV) gap in density of states at the antinodes~\cite{ding1, damascelli1, hossain1}.

An unambiguous interpretation of quantum oscillation results in terms of charge carrier type, and single or multiple surfaces in the case of the underdoped cuprates has thus far been prone to pitfalls, due to the quasi-two-dimensionality of the Fermi surface, closeness of frequencies and effective cyclotron masses of the low frequency components, and the lack of clarity regarding the observation of a significantly larger frequency~\cite{sebastian2, audouard1, sebastian3, ramshaw2}. Quantum oscillations in the case of a fixed chemical potential cannot discern the sign of charge carriers or whether there is a single or multiple types of charge carrier present~\cite{shoenberg2}, causing recourse thus far to less direct probes such as the Hall effect and thermopower, and calculations of the electronic structure to infer a Fermi surface comprising mobile electrons and less mobile holes~\cite{leboeuf1, rourke1, leboeuf2, lilberte1, sebastian3}. 

In this work, we observe well-defined second harmonic quantum oscillations from a broad array of torque and contactless resistivity experiments performed in a 45T DC magnet, a 60T generator-driven magnet, and 65 $\&$ 85T pulsed magnets in angles 0 $\lessapprox \theta \lessapprox$ 60$^\circ$ and temperatures 1K $\lessapprox T \lessapprox$ 5 K in the National High Magnetic Field Laboratory. From the analysis of the measured second harmonic oscillations, we present the surprising finding that the harmonic content in underdoped YBa$_2$Cu$_3$O$_{6+x}$ ($x=$~0.56) arises from oscillations of the chemical potential, enhanced compared to that in normal metals by the quasi-two-dimensional topology of the Fermi surface~\cite{shoenberg2, shoenberg1,harrison1, usher1, yoshida1}. Oscillatory features corresponding to multiple frequencies in $1/B$ previously reported~\cite{doiron1,audouard1,sebastian2,sebastian3} are observed: ($F_\alpha=$ 535(5)T, $F_{\gamma 1}=$ 440(10)T, $F_{\gamma 2}=$ 620(20)T) down to 22T, and $F_\beta=$ $1650(20)T$ at the lowest temperatures~\cite{sebastian5}; yet we conclude from the magnitude and phase of the second harmonic of the chemical potential with respect to the fundamental oscillations, that the Fermi surface consists only of a single quasi-two-dimensional pocket, and no other Fermi surface section of significant thermodynamic mass, the multiple frequency components arising from effects of finite $c$-axis dispersion, bilayer splitting, and magnetic breakdown. Given the nodal concentration of the density of states observed in zero field photoemission~\cite{ding1, damascelli1, hossain1} and high field heat capacity experiments~\cite{riggs1}, indications are that, in the absence of a significant effect of magnetic field, the observed Fermi surface pocket is located at the nodal region of the Brillouin zone. This finding does not, however, necessarily imply a contradiction with the observation of a negative Hall and Seebeck coefficient~\cite{leboeuf1,sebastian5}. 

\section{Results}
\begin{figure*}[htbp!]
\centering
\includegraphics[width=0.8\textwidth]{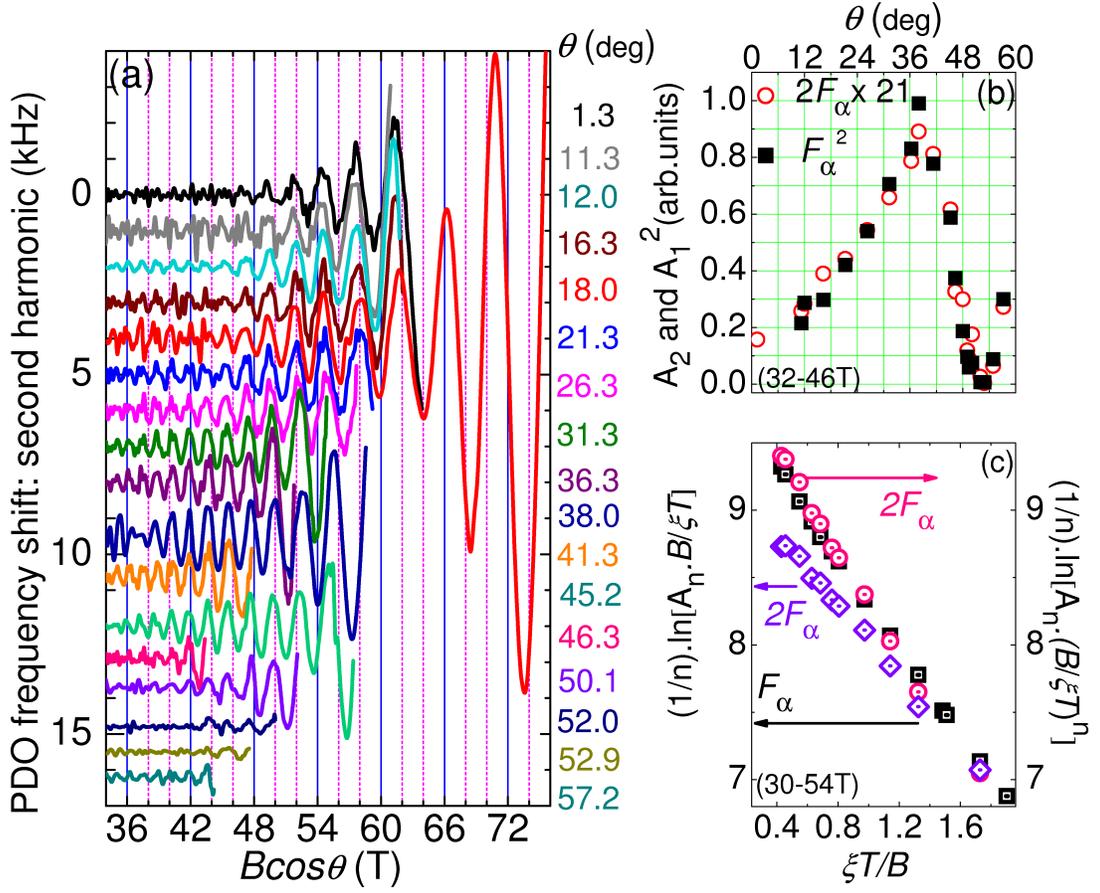}
\caption{
{\bf a}, Field-dependence of the second harmonic (2$F$) from the data in Ref.~\cite{sebastian5} at different angles $\theta$ between the crystalline c-axis and the magnetic field after having subtracted fits to the fundamental ($F$) and a slowly varying background. The data are plotted versus $B\cos\theta$ to facilitate a comparison of the relative phase. The second harmonic amplitude falls to a small value near 54$^\circ$ (at which the fundamental undergoes a spin zero~\cite{sebastian5}), following which it recovers with the same phase at higher angles. {\bf b}, A comparison of the Fourier amplitude of the second harmonic obtained over the field interval indicated against the rescaled square of the amplitude of the fundamental ($F^2$) reveals that they precisely track each other over a broad range of angles, spanning the spin zero in the fundamental near 54$^\circ$. {\bf c}, Temperature-dependence of the fundamental (black squares) and second harmonic (colored squares). We plot $(1/n)\ln[A_n B/\xi T]$ versus $\xi T/B$, where $n$ is the harmonic index, $\xi = 2\pi^2 k_{\rm B}m_{\rm e}/\hbar e = 14.69$T/K and $A_n$ is the measured harmonic amplitude. The slope of this relation is expected to have the value $\dfrac{m^\ast}{m_\textrm e}$, where $m_{\rm e}$ is the free electron mass, for both the second harmonic and the fundamental (given the division by $n$ on the left axis). We find, however, an unexpectedly low slope of value 1.3(1) for the second harmonic, which is lower than the slope $\dfrac{m^\ast}{m_e} = 1.6(1)$ measured for the fundamental. On plotting $(1/n)\ln[A_n(B/\xi T)^n]$ versus $\xi T/B$ for the second harmonic (pink symbols, right axis), however, the slope of the relation agrees with that of the fundamental. The latter implies that the temperature dependence of the second harmonic is given by $R_{T1}^2$ (where $R_{T1}=X/\sinh X \approx 2 X e^{-X}$ for $X=\xi Tm^\ast/Bm_{\rm e} \ll 1$), as expected from the effects of oscillations in the chemical potential~\cite{gold1}.
}
\label{angletemperature}
\end{figure*}

A number of observations relating to the second harmonic reveal departures from the oscillatory behavior expected from the standard Lifshitz-Kosevich (LK) theory that assumes a fixed chemical potential~\cite{shoenberg2}. Figure~\ref{angletemperature}{\bf a} shows the second harmonic extracted from quantum oscillations measured on YBa$_2$Cu$_3$O$_{6+\textrm x}$ at a series of fixed angles $\theta$ (between the crystalline $c$-axis and the magnetic field) as a function of the $c$-axis magnetic field component $B\cos\theta$, where $B=\mu_0H$ is the applied magnetic field (up to 65T in a pulsed field magnet). We find that the second harmonic oscillations retain a constant phase between $\theta=$~0$^\circ$ and 57$^\circ$ at fixed $B\cos\theta$~\cite{sebastian3}. A remarkable feature in the data is the absence of a zero near $\approx$~18$^\circ$, which is expected in the second harmonic if $\mu$ is fixed, given the presence of a spin zero in the fundamental at $\approx$~54$^\circ$~\cite{sebastian5} (see Fig.~\ref{angletemperature}{\bf b}).  Instead, the second harmonic amplitude falls to zero at precisely the same angle as the fundamental.

Figure~\ref{angletemperature}{\bf c} compares the $T$ dependence of the fundamental amplitude $A_1$, with that of the second harmonic amplitude $A_2$ on a linear-log plot normally used to infer the appropriate cyclotron mass. These precisely temperature controlled measurements were performed with the sample in liquid $^4 \textrm {He}$ in a 60T generator-driven magnet. Quantum oscillations corresponding to a fixed chemical potential are expected to exhibit a second harmonic with a cyclotron mass twice the fundamental cyclotron mass, yet strangely a fit of the second harmonic to the thermal damping factor (see Fig.~\ref{angletemperature}{\bf c}) yields a mass that is too low by about 25 $\%$.

\begin{figure*}[htbp!]
\centering
\includegraphics[width=0.75\textwidth]{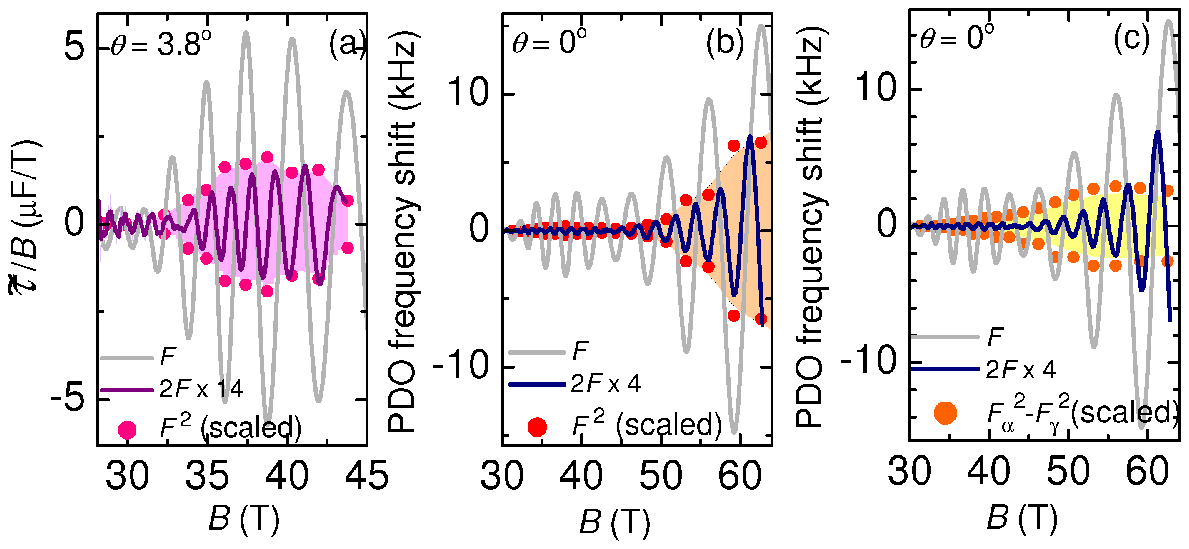}
\caption{{\bf a}, A comparison of the second harmonic oscillations (thick purple lines) with the scaled square of the amplitude of the fundamental (dots) versus magnetic field, indicating that they are closely correlated for the oscillatory magnetization $\tilde{M}\propto \tilde{\tau}/B$ extracted from the magnetic torque $\tilde{\tau}$ at an angle $\theta\approx$~3.8$^\circ$ measured in the 45T DC magnet. The dots are extracted from the scaled square of the fundamental oscillations maxima and minima.
The actual fundamental oscillations are depicted in light grey. {\bf b},  The same repeated for the oscillatory contactless resistivity data at $\theta\approx$~0 measured in the 65T pulsed magnet, showing the same behavior as in {\bf a}.  {\bf c}, A comparison of the second harmonic oscillations (thick navy lines) with the scaled difference between the squared amplitude of the fundamental $\alpha$ and $\gamma$ components ($\alpha$ and $\gamma$ component fit from ref.~\cite{sebastian3}). This would be the expectation for $\alpha$ and $\gamma$ corresponding to opposite carrier types, we find that the observed second harmonic does not follow this form.
}
\label{square}
\end{figure*}

Figure~\ref{square} shows the detected amplitude of the second harmonic as a function of magnetic field measured using two different methods: magnetic torque in DC fields and contactless resistivity in pulsed fields. The measured second harmonic amplitude is found to scale precisely as the square of the fundamental as a function of magnetic field. This square relationship mirrors that as a function of angle (Fig.~\ref{angletemperature}{\bf b}) and as a function of temperature (Fig.~\ref{angletemperature}{\bf c}).

\begin{figure*}[htbp!]
\centering
\includegraphics[width=0.75\textwidth]{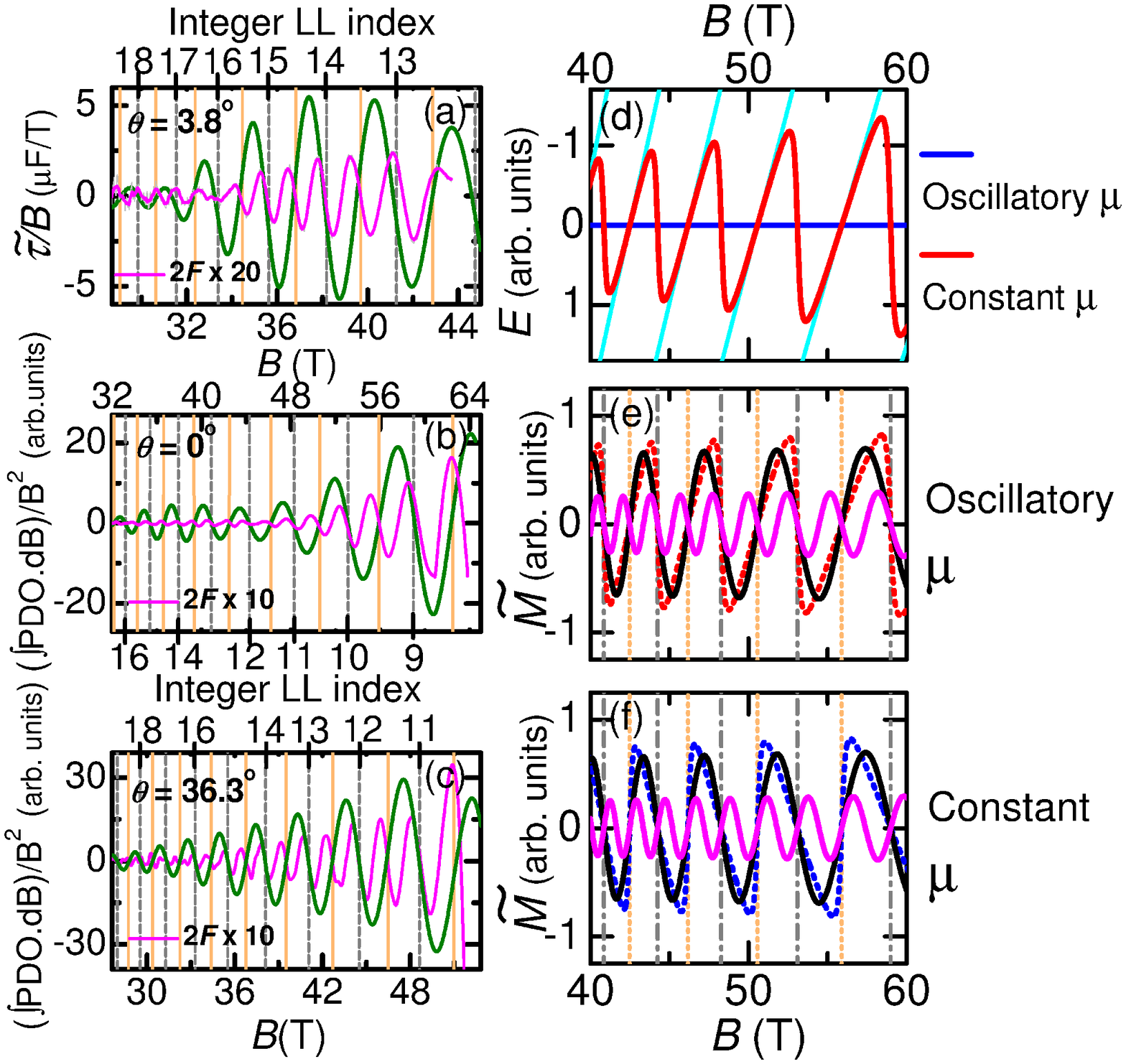}
\caption{
{\bf a} - {\bf c}, The fundamental (green) and second harmonic (magenta) extracted from the magnetic torque measured in YBa$_2$Cu$_3$O$_{6+x}$ at different field orientations $\theta$ as indicated, revealing the same phase behavior as seen in the schematic figure ({\bf e}). The magnetization is inferred using $\tilde{M}\propto\tilde{\tau}/B$ and $\tilde{M}\propto\int\tilde{\rho}{\rm d}B/B^2$ from the magnetic torque $\tilde{\tau}$ and contactless resistivity respectively. Integer Landau level filling factors (when $\mu$ resides in the gap between two-dimensional $\alpha$ component Landau levels) are indicated.
{\bf d}, A schematic showing the Landau level energies (cyan lines) and how the chemical potential $\mu$ (red line) and consequently the magnetization (red line in {\bf e}) oscillates for a single two-dimensional electron pocket. $\mu$ jumps between Landau levels (cyan lines) at integer filling factors causing it to be momentarily situated in the Landau gap. {\bf e}. The magnetization (red) together with the extracted fundamental (black) and second harmonic (magenta). {\bf f}, The corresponding magnetization waveform (blue) when the chemical potential is fixed, giving rise to a second harmonic of opposite sign. Vertical lines in gray correspond to integral Laundau levels, and in orange correspond to half-integral Landau levels.
}
\label{sawtooth}
\end{figure*}

Fig.~\ref{sawtooth} allows us to infer the phase relation between the observed second harmonic oscillations and the fundamental oscillations at 3.8$^\circ$ in the measured torque, 0$^\circ$ in the contactless resistivity signal and at 38$^\circ$ in the contactless resistivity. The second harmonic retains a constant phase with respect to the fundamental, consistent with the sawtooth waveform shown schematically in Figs.~\ref{sawtooth}{\bf d} and {\bf e} respectively. This is inconsistent with that expected for quantum oscillations corresponding to a fixed chemical potential, which leads to a different form as shown in Fig.~\ref{sawtooth}{\bf f}) and to a spin zero and phase inversion at $18^\circ$, which we do not observe.

Finally, Fig.~\ref{scalings}a shows a collation of quantum oscillations measured in magnetic torque and contactless resistivity, spanning overall a range 22T to 65T. The identical beat structure previously revealed to correspond to three closely spaced frequencies ($F_\alpha=$ 535(5)T, $F_{\gamma 1}=$ 440(10)T, $F_{\gamma 2}=$ 620(20)T) is found to persist down to 22T, confirming the multiple frequency components earlier reported~\cite{doiron1,audouard1,sebastian2,sebastian3,beta1}.

\section{Discussion}

We broadly consider three chief aspects of the experimental data, and understand the unusual behaviour of the second harmonic in terms of the effect of oscillations of the chemical potential $\mu$. 

Large oscillations in $\mu$ occur in layered metals owing to the absence of a large reservoir of  phase-incoherent states to pin $\mu$ at an essentially constant value~\cite{harrison1, usher1, shoenberg1}. Instead of the Landau levels passing through a constant chemical potential $-$ as described in the simplest LK theory~\cite{shoenberg2} $-$ $\mu$ itself oscillates. The chemical potential remains pinned to the highest partially occupied Landau level as its degeneracy changes, jumping from one to the next as the levels become sequentially depopulated on increasing the magnetic field (shown schematically in Fig.~\ref{sawtooth}{\bf d}). Oscillations in the magnetization (the de~Haas-van~Alphen effect) are affected by the jumps in $\mu$, yielding a waveform that resembles a sawtooth (see schematic in Fig.~\ref{sawtooth}{\bf e}). 

We consider the effect that the oscillatory component $\tilde{\mu}$ has on modulating the Onsager phase of the quantum oscillations $\phi=2\pi F/B-\eta$ (where $\eta\approx\pi$) arising from a single two-dimensional pocket giving rise to quantum oscillations of frequency $F$~\cite{shoenberg2}.  The oscillatory magnetization is given by
\begin {eqnarray}
\dfrac{\tilde{M}}{M_0} \propto \big( \kappa_{1}\sin \phi + \frac{\kappa_{2}}{2}\sin 2\phi + \dots \big)
\label{magnetization1}
\end{eqnarray}
where $\kappa_{p}$ is the product of quantum oscillation amplitude damping factors (e.g. Dingle, thermal, spin, mosaic spread, etc.) affecting the $p^{\rm th}$ harmonic~\cite{shoenberg2}, and $M_0$ is the absolute amplitude at infinite field. The effect of chemical potential oscillations are explicitly included in leading order by the expansion of $\phi(\mu)=\phi_0+\tilde{\mu}(\partial\phi_0/\partial\bar{\mu})$, where $\phi_0=\phi(\bar{\mu})$, and $\bar{\mu}$ is the steady part of $\mu$~\cite{gold1,methods}. This allows us to rewrite Eqn.~\ref{magnetization1} in the form
\begin {eqnarray}
\dfrac{\tilde{M}}{M_0} \propto \big( \kappa_{1}\sin\phi_0-\frac{\small 1}{\small 2}\big(2\Gamma \kappa_{1}^2-\kappa_{2}\big)\sin 2\phi_0+\dots \big)
\label{magnetization2}
\end{eqnarray}
where $\Gamma$ is determined by the relative strength of $\tilde{\mu}$ as discussed below~(\cite{gold1,shoenberg2}). While the fundamental frequency continues to be well described by the standard theory for oscillations corresponding to fixed $\mu$, the second harmonic can be dominated by a new contribution of relative weight $\dfrac{2\Gamma\kappa_1^2}{\kappa_2}$ (see Eqn.~\ref{magnetization2}) resulting from the oscillations in $\mu$. Here $\Gamma$ represents the fraction of the total density of states associated with the postulated two-dimensional pocket.

\begin{figure*}[htbp!]
\centering
\includegraphics[width=0.75\textwidth]{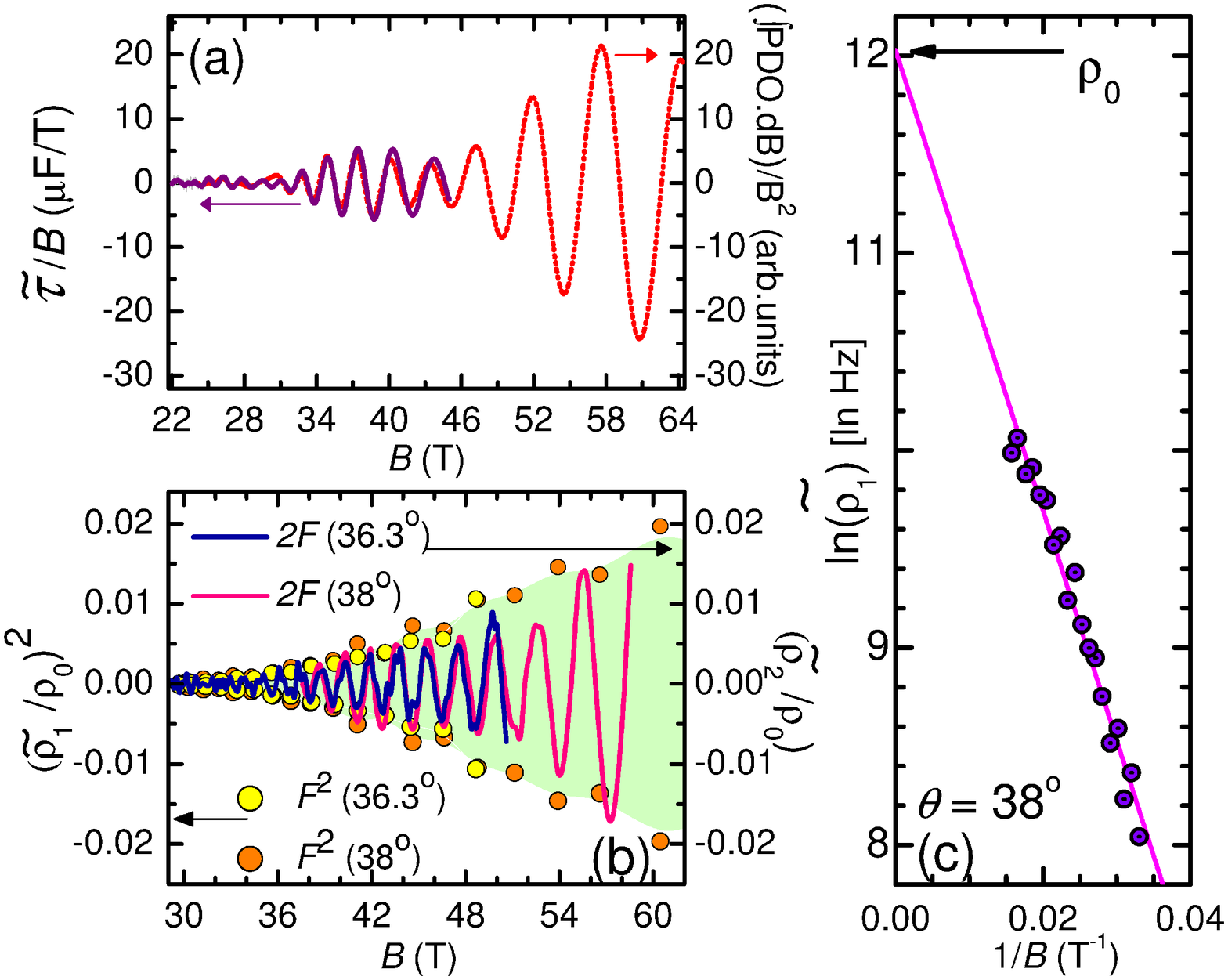}
\caption{{\bf a}, Comparison of the magnetization $\tilde{M}\propto\tilde{\tau}/B$ and $\tilde{M}\propto\int\tilde{\rho}{\rm d}B/B^2$ inferred from the measured magnetic torque $\tilde{\tau}$ (purple line) and contactless resistivity $\tilde{\rho}$ (red dotted line) respectively. 
{\bf b}, Plot of the second harmonic (at $\theta\approx$~36.3$^\circ$ in fields to 65~T and at $\theta\approx$~38$^\circ$ in fields 85~T) divided by the infinite field intercept $\rho_0$ determined in ({\bf c}): $(\tilde{\rho}_2/\rho_0)$ represented by solid lines is compared with $(\tilde{\rho}_1/\rho_0)^2$, represented by solid dots. Here the amplitude of the fundamental ($\tilde{\rho}_1$) is determined in the same manner as in Fig.~2. The ratio of the amplitude of the harmonic (solid lines) to the square of the fundamental (dots) yields a value of 2$\Gamma_{38^\circ}=$~1.0(3) from Eqn.~\ref{magnetization2}, from which we obtain $\Gamma_{38^\circ}=$~0.50(15).
{\bf c}, Dingle analysis, performed on a linear-log plot of the fundamental oscillations in contactless resistivity (denoted by solid circles, determined from the oscillation maxima and minima after dividing out $R_{T,1}$) at $\theta=$~38$^\circ$ up to fields of 85~T, plotted versus $1/B$. The $1/B=0$ intercept obtained by fitting a straight line (purple) to this plot yields $\rho_0=$~(1.6(5)~$\times$~10$^5)$~Hz.
}
\label{scalings}
\end{figure*}

In Fig.~\ref{scalings}{\bf a} we compare the fundamental and second harmonic magnetization inferred from the  measured torque $\tilde{M}\propto\tilde{\tau}/B$ and from the contactless resistivity $\tilde{M}\propto\int\tilde{\rho}{\rm d}B/B^2$~\cite{shoenberg2}, where the integration is over the experimental field range. Their similar behavior and scalability indicates that the oscillations in both quantities have a closely related origin~\cite{shoenberg2}.

First we discuss the phase of the observed second harmonic oscillations. We see that in the case of strong chemical potential oscillations (where the first term in the prefactor of $\sin 2\phi_0$ in Eqn.~\ref{magnetization2} is dominant), the second harmonic oscillations are in fact out of phase by $\pi$ from the conventional case of oscillations corresponding to a fixed chemical potential $-$ precisely as we observe in experiment (predictions and comparison with experiment shown in Fig.~\ref{sawtooth}d.)

Secondly, we discuss the amplitude $A_2$ of the second harmonic, which is expected to vary as the square of the amplitude $A_1$ of the fundamental in the limit 2$\Gamma \kappa_1^2 > \kappa_2$ in the case where harmonic enrichment arises mainly from chemical potential oscillations. The experimentally observed square relation $A_2 \propto A_1^2$ as a function of magnetic field at various angles both in the torque measurements and contactless conductivity (Fig.~\ref{square} and Fig.~\ref{scalings}b) is consistent with this expectation. Also consistent with this expectation is the observed absence of a low-angle spin zero as discussed earlier for the second harmonic (Fig.~\ref{angletemperature}b( and the low value of the effective mass inferred from a conventional linear-log plot of the second harmonic as also mentioned earlier (Fig.~\ref{angletemperature}c).

Thus far we have considered a model with a single two-dimensional Fermi surface pocket and hence quantum oscillations with a single periodicity in $1/B$. The predictions of a model including multiple frequencies arising from warping and bilayer splitting of the same pocket are not substantially different from those discussed above. However, the inclusion of a second two-dimensional Fermi surface pocket associated with carriers of opposite effective charge would lead to an additional harmonic enrichment term from chemical potential oscillations of the opposite sign (i.e.) $-\kappa_1'^2$ in addition to $-\kappa_1^2$ in Eqn.~\ref{magnetization2}, leading to behaviour in general inconsistent with harmonic components scaling as the square of fundamental components~\cite{shoenberg2}. Fig.~\ref{square}c shows the predicted behaviour for the second harmonic were the $\alpha$ and $\gamma$ component of the oscillations to be associated with pockets of opposite carrier type $-$ this does not fit the experimentally observed form of the second harmonic, ruling out the possibility of the observed quantum oscillations arising from pockets containing opposite carriers. 

Our finding $A_2 \propto A_1^2$ therefore supports the idea that the observed frequency components are all closely related and are likely to arise from the same two-dimensional pocket, its purely two-dimensional structure modulated by weak warping and bilayer splitting. We next consider what can be learned from the magnitude of the ratio of $A_2$ to $A_1^2$, namely, from the magnitude of $\Gamma$ as defined in Eqn.~\ref{magnetization2}. This ratio is most accurately determined at $\theta=38^\circ$ where the $\gamma$ oscillatory component is weak and the quantum oscillations are dominated by a single periodicity due to the $\alpha$ component~\cite{sebastian3,ramshaw2,sebastian5}. In this case $\Gamma \approx \dfrac{\textrm{DOS}(\alpha)}{\textrm{DOS}(\alpha) + \textrm{DOS}(\gamma) + \textrm{DOS} (\textrm{res})}$, where `DOS' stands for density of states and `res' for any reservoir of carriers at the chemical potential not associated with the $\alpha$ or $\gamma$ components. From the measured value $\Gamma(38^\circ) = 0.50(15)$ at this angle (see Fig.~\ref{scalings}b), we estimate an upper bound for $\dfrac{\textrm{DOS(res)}}{\textrm{DOS}(\alpha) + \textrm{DOS}(\gamma)}$ of about $30\%$. This suggests that our model of a single quasi-two-dimensional Fermi surface pocket, giving rise to multiple frequency components $\alpha$ and $\gamma$ together with possible magnetic breakdown components, provide an essentially complete description of the Fermi surface under our experimental conditions in underdoped YBa$_2$Cu$_3$O$_{6+\textrm x}$.

Complementary photoemission~\cite{ding1, damascelli1, hossain1} and high field heat capacity~\cite{riggs1} experiments yield a concentration of density of states at the nodal location in momentum space, suggesting that in the absence of a significant effect of magnetic field, the single Fermi surface pocket that gives rise to multiple frequency components observed by quantum oscillations in underdoped YBa$_2$Cu$_3$O$_{6+x}$ is also connected with the nodal region of the Brillouin zone.

\section{conclusion}
In summary, we find that second harmonic quantum oscillations in underdoped YBa$_2$Cu$_3$O$_{6+\textrm x}$ arise mainly from the effect of oscillations of the chemical potential. An analysis of the phase and magnitude of the second harmonic components relative to the fundamental components allows us to conclude that the Fermi surface consists of a single quasi-two-dimensional pocket, the multiplicity of frequency components observed being a consequence of warping, bilayer splitting along with the effects of magnetic breakdown that can produce extended orbits, and also potentially open orbits. This pocket is likely located at the nodal region, since photoemission~\cite{ding1, damascelli1, hossain1} and high field heat capacity~\cite{riggs1} experiments find a concentration in the density of states at the nodes of the superconducting wavefunction. Given that within translational symmetry breaking models proposed thus far (e.g.~\cite{millis1, chakravarty1}), the nodal pockets comprise hole carriers, we are faced with a dilemma as to how to explain the high field quantum oscillations with a negative Hall and Seebeck effect~\cite{leboeuf1, lilberte1} in terms of these models, unless the Hall and Seebeck effects do not reflect the sign of carrier in the Fermi surface pockets, or unless a significant change in electronic structure is induced by a magnetic field. Possible resolutions to this problem involve understanding how a negative Hall and Seebeck effect can arise from a nodal pocket~\cite{sebastian5}.


\begin{thebibliography}{99}

\bibitem{bednorz1} Bednorz,~J.~G., Muller, K.~A. Possible high-$T_{\rm c}$ superconductivity in the Ba-La-Cu-O system. {\it Z. Phys. B} {\bf 64}, 189-193 (1986).

\bibitem{bonn1} Bonn,~D.~A. Are high-temperature superconductors exotic? {\it Nat. Phys.} {\bf 2}, 159-168 (2006).

\bibitem{doiron1} 
Doiron-Leyraud, N. {\it et al.} 
Quantum oscillations and the Fermi surface in an underdoped high -$T_{\rm c}$ superconductor {\it Nature} {\bf 447}, 565-568 (2007).

\bibitem{leboeuf1}
LeBoeuf, D. {\it et al.} Electron pockets in the Fermi surface of holed-doped high $T_{\rm c}$ superconductors. 
{\it Nature} {\bf 450}, 533-536 (2007).

\bibitem{jaudet1} Jaudet,~C. {\it et al.} De~Haas-van~Alphen oscillations in the underdoped high-temperature superconductor YBa$_2$Cu$_3$O$_{6.5}$. {\it Phys. Rev. Lett.} {\bf 100}, 187005 (2008).

\bibitem{sebastian2}
Sebastian, S.~E. {\it et al.} A multi-component Fermi surface in the vortex state of an underdoped high $T_{\rm c}$ superconductor.
{\it Nature} {\bf 454}, 200-203 (2008).

\bibitem{audouard1} Audouard, A. {\it et al.} Multiple quantum oscillations in the de~Haas-van~Alphen spectra of the underdoped high-temperature superconductor YBa$_2$Cu$_3$O$_{6.5}$.
{\it Phys. Rev. Lett.} {\bf 103}, 157003 (2009).

\bibitem{sebastian1}
Sebastian, S.~E. {\it et al.} Fermi-liquid behavior in an underdoped high-$T_{\rm c}$ superconductor.
{\it Phys. Rev. B} {\bf 81}, 140505 (2010).

\bibitem{sebastian3} Sebastian, S.~E. {\it et al.} Compensated electron and hole pockets in an underdoped high-$T_{\rm c}$ superconductor.
{\it Phys. Rev. B} {\bf 81}, 214524 (2010).

\bibitem{ramshaw2} Ramshaw,~B.~J. {\it et al.} 
{\it Nat. Phys.} (in press 2011).

\bibitem{millis1}
Millis, A.~J. and Norman, M.~R. Antiphase stripe order as the origin of electron pockets observed in $1/8$-hole-doped cuprates. 
{\it Phys. Rev. B} {\bf 76}, 220503 (2007).

\bibitem{chakravarty1} Chakravarty,~S., .Kee,~H.-Y. Fermi pockets and quantum oscillations of the Hall coefficient in high-temperature superconductors. 
{\it Proc. Nat. Acad. Sci. USA} {\bf 105}, 8835-8839 (2008).

\bibitem{lee1} Lee, P.~A., Doping a Mott insulator: Physics of high-temperature superconductivity. {\it Rep. Prog. Phys.} {\bf 71}, 012501 (2008).

\bibitem{ding1} Ding, H. {\it et al.},
Spectroscopic evidence for a pseudogap in the normal state of underdoped high-T-c superconductors.
{\it Nature} {\bf 382}, 51-54 (1996).

\bibitem{damascelli1}
Damascelli, A., Hussain, Z., Shen, Z.~X., 
Angle-resolved photoemission studies of the cuprate superconductors. 
{\it Rev. Mod. Phys.} {\bf 75}, 473-541 (2003).

\bibitem{hossain1} Hossain,M.~A. {\it et al.} In situ doping control of the surface of high-temperature superconductors 
{\it Nature Phys.} {\bf 4}, 527-531 (2008).

\bibitem{riggs1} Riggs, S.~C. {\it et al.}, 
Planes, Chains, and Orbits: Quantum Oscillations and High Magnetic Field Heat Capacity in Underdoped YBCO
{\it Nat. Phys.} (in press 2011).

\bibitem{beta1} The significantly higher $F_{\beta} \approx 1650(50)$T frequency is also prominently observed in quantum oscillations measured at the lowest temperatures.

\bibitem{shoenberg2} Shoenberg, D., {\it Magnetic oscillations in metals} (Cambridge University Press,
Cambridge 1984).

\bibitem{rourke1} Rourke, P.~M.~C.~{\it et al.} 
Fermi-surface reconstruction and two-carrier model for the Hall effect in YBa$_2$Cu$_4$O$_8$. 
{\it Phys. Rev. B} {\bf 82}, 020514 (2010).

\bibitem{leboeuf2} LeBoeuf, D. {\it et al.}, 
Lifshitz critical point in the cuprate superconductor YBa$_2$Cu$_3$O$_y$ from high-field Hall effect measurements. 
{\it Phys. Rev. B} {\bf 83}, 054506 (2011).

\bibitem{lilberte1} Laliberte, F. {\it et al.} Fermi-surface reconstruction by stripe order in cuprate superconductors
preprint arXiv:1102.0984.

\bibitem{shoenberg1} Shoenberg, D., Magnetization of a two-dimensional electron-gas. {\it J. Low. Temp. Phys.} {\bf 56}, 417-440 (1984).

\bibitem{harrison1} Harrison, N.~{\it et al.} 
Numerical model of quantum oscillations in quasi-two-dimensional organic metals in high magnetic fields.
{\it Phys. Rev. B} {\bf 54}, 9977-9987 (1996).

\bibitem{usher1} Usher, A., Elliott, M., Magnetometry of low-dimensional electron and hole systems. 
{\it J. Phys.-Cond. Matt.} {\bf 21}, 103202 (2009).

\bibitem{yoshida1} Yoshida, Y.~{\it et al.} 
Fermi surface properties in Sr$_2$RuO$_4$.  
{\it J. Phys. Soc. Japan} {\bf 68}, 3041-3053 (1999).

\bibitem{sebastian5} Sebastian, S.~E. {\it et al.} (arXiv.1103.xxxx).

\bibitem{gold1} Gold, A.~V. The De Haas-Van Alphen Effect in Solid State Physics ed. J. F. Cochran and R. R. Haering (New York: Gordon and Breach) vol 1 p 39 (1968).

\bibitem{methods} The $\sin[p\phi(\mu)]$ terms are expanded after having made the substitution $\tilde{\mu}(\partial\phi_0/\partial\bar{\mu}) \propto \sin[\phi_0]$.

\end{thebibliography}
\end{document}